# Viscous flow properties and hydrodynamic diameter of phenothiazine-based redox-active molecules in different supporting salt environments

Yilin Wang[a,b,c], Aman Preet Kaur[d,e], N. Harsha Attanayake[d,e], Zhou Yu[f], Thilini M. Suduwella[d,e], Lei Cheng[f], Susan A. Odom[d,e], and Randy H. Ewoldt[*,a,b,c,g]

[a] *Department of Mechanical Science and Engineering, University of Illinois at Urbana-Champaign, Urbana, IL 61820 USA*
[b] *Beckman Institute for Advanced Science and Technology, University of Illinois at Urbana-Champaign, Urbana, IL 61820 USA*
[c] *Joint Center for Energy Storage Research, Urbana, IL 61820 USA*
[d] *Joint Center for Energy Storage Research, Lexington, KY 40506 USA*
[e] *Department of Chemistry, University of Kentucky, Lexington, KY 40506, USA*
[f] *Materials Science Division and Joint Center for Energy Storage Research, Argonne National Laboratory, Lemont, IL 60439 USA*
[g] *Materials Research Laboratory, University of Illinois at Urbana-Champaign, Urbana, IL 61820 USA*

## ABSTRACT

We report viscous flow properties of a redox-active organic molecule, *N*-(2-(2-methoxyethoxy)ethyl)phenothiazine (MEEPT), a candidate for non-aqueous redox flow batteries, and two of its radical cation salts. A microfluidic viscometer enabled the use of small sample volumes in determining viscosity as a function of shear rate and concentration in the non-aqueous solvent, acetonitrile, both with and without supporting salts. All solutions tested show Newtonian behavior over shear rates of up to 30,000 s$^{-1}$, which is rationalized by scaling arguments for the diffusion-based relaxation time of a single MEEPT molecule without aggregation. Neat MEEPT is flowable but with a large viscosity (412 mPa·s at room temperature), which is approximately 1,000 times larger than acetonitrile. When dissolved in acetonitrile, MEEPT solutions have low viscosities; at concentrations up to 0.5 M, the viscosity increases by less than a factor of two. From concentration-dependent viscosity measurements, molecular information is inferred from intrinsic viscosity (hydrodynamic diameter) and the Huggins coefficient (interactions). Model fit credibility is assessed using the Bayesian Information Criterion (BIC). It is found that the MEEPT and its charged cation are "flowable" and do not flocculate at concentrations up to 0.5 M. MEEPT has a hydrodynamic diameter of around 8.5 Å, which is largely insensitive to supporting salt and state of charge. This size is comparable to molecular dimensions of single molecules obtained from optimized structures using density function theory calculations. The results suggest that MEEPT is a promising candidate for redox flow batteries in terms of its viscous flow properties.

# 1. INTRODUCTION

Non-aqueous redox flow batteries (NAqRFBs) utilizing solutions of redox-active organic molecules (ROMs) are competitive for reliable electrochemical energy storage systems due to their scalable energy capacity, large electrochemical stability windows, and potentially long operating lifetimes[1–6]. In NAqRFBs, charges are stored in ROMs and transported by ionic supporting salts, both are dissolved in an organic solvent. From a techno-economic study,[2,7] the concentration of active materials must be larger than 1 M for NAqRFBs to be competitive, with a target concentration of 5 M. Previous research has shown that highly concentrated electrolytes can result in dramatically increased viscosity.[8] The performance of RFBs is greatly related to the viscosity of electrolytes, with higher viscosities having a prominent negative influence on key transport properties[9], such as ionic conductivity[10] and diffusivity.[11,12] Viscosity is also directly related to pumping costs.[13] Bindner *et al.* reported an 8-11% of total power loss from pumping for a vanadium RFB.[14] Understanding the origin of viscosity differences in these complex fluids is therefore of interest in the development of solutions with favorable properties.

Studies evaluating the viscous flow properties of electrolytes for RFBs show that the concentration-dependent viscosities of electrolytes can be affected by ROM size[15,16] and its state of charge,[17,18] solution temperature,[19] inclusion of additives,[20] and ionic strength of supporting salts.[21] Though efforts have been made in predicting electrolyte viscosity,[9,22–24] no universal method accounts for all of these factors in the absence of experimental calibration. Furthermore, few publications discuss non-Newtonian behavior of ROM electrolytes, which might arise from flow-induced conformation of ROMs[25] and the break-up of interactions between ROMs.[26] Non-Newtonian analysis matters in non-equilibrium molecular dynamics (NEMD) simulations,[27–31] as the NEMD simulation may study flows in the large Weissenberg number (Wi) regime (very high strain rates), where viscosities show non-Newtonian behavior. Therefore, to optimize the operating condition of RFBs, experimental viscous flow properties of ROM solutions must be considered.

In this study, the viscous flow properties of *N*-(2-(2-methoxyethoxy)ethyl)phenothiazine (MEEPT) were studied (Figure 1). MEEPT is a ROM that shows promise characteristics for meeting requirements for grid-scale energy storage, as evidenced by its high current density and long duration cycling.[32] However, its viscosity and other transport properties, such as diffusivity and ionic conductivity, have not been reported. Here we measured the concentration-dependent viscosities of neutral MEEPT, its tetrafluoroborate radical-cation salt (MEEPT-BF$_4$), and its bis(trifluoromethanesulfonyl)imide salt (MEEPT-TFSI) at different concentrations in acetonitrile (ACN) and in ACN-based electrolytes containing different supporting salts, namely, tetraethylammonium tetrafluoroborate (TEABF$_4$) and lithium bis(trifluoromethanesulfonyl)imide (LiTFSI). For concentration-dependent viscosities, the Einstein,[33] Huggins,[34] and extended Jones-Dole equations[35] were used to fit and understand the measured viscosities. From fitting results, molecular information, such as intrinsic viscosity, hydrodynamic diameter, Huggins coefficient, and the presence of ion interactions can be determined. The residual sums of squares (RSS) were used to fit and the Bayesian Information Criteria (BIC) of fits were compared to determine the most credible fit. A simple estimate of the relaxation time scale of MEEPT was conducted to understand the observed Newtonian behavior of MEEPT solutions.

The hydrodynamic diameter of MEEPT remained essentially unchanged with the addition of supporting salts, and was comparable to the molecular structure dimensions, which suggests a minimal flocculation and small solvation shells in supporting salt environments. The Huggins coefficients were of the same order of magnitude, but larger than that of Brownian hard spheres in shear flow, derived by Batchelor,[36] suggesting that interactions other than hydrodynamic interactions and steric effects contribute to viscous dissipation. Our results ultimately show that (i) both neutral and charged MEEPT are "flowable" ROMs, and their solution viscosities increase with concentration as a result of excluded volume and interactions between ROMs, (ii) addition of supporting salts does not have a dramatic influence on the hydrodynamic diameter of MEEPT, but does affect the interactions between MEEPT molecules, (iii) the viscosities increase in the charged form of MEEPT, which arise from the volume excluded by corresponding anions, and (iv) there is minimal flocculation within the concentrations and shear rates tested, and the critical shear rate, the rate at which the solution begins to show shear-thinning, is larger than limits experimentally accessed.

## 2. MATERIALS AND METHODS

### A. Materials

The redox-active molecule MEEPT (301.40 g/mol) was obtained from TCI, and its radical cation salts, MEEPT-BF$_4$[32] and MEEPT-TFSI[37] were synthesized as described in the Supporting Information. Two supporting salts, lithium bis(trifluoromethane)sulfonimide (LiTFSI, Aldrich, 99.95%) and tetraethylammonium tetrafluoroborate (TEABF$_4$, Aldrich, 99%), were chosen because these are both commonly used in RFBs and they affect the solubility of charged MEEPT differently, as shown in Table 1. All ROMs and salts were dissolved in acetonitrile (ACN) to reduce ion associations, as ACN is the most polar organic solvent with a wide electrochemical stability window,[38,39] and therefore, it is commonly employed in NAqRFBs.

### B. Viscometry

The dynamic viscosity of solutions was measured using a microfluidic viscometer m-VROC (RheoSense Inc.). Only microliter sample volumes are required by this viscometer, and free surface effects,[40] evaporation, and contamination can be avoided during the measurement due to internal flow. The measuring chip is made from borosilicate glass containing a rectangular slit flow channel with uniform cross-sectional area, as shown in Figure 1(a).[41] The fluid sample is pushed by a syringe pump at a constant volume flow rate, $Q$, which is related to the apparent shear rate, $\dot{\gamma}_{app}$, by[25]

$$\dot{\gamma}_{app} = \frac{6Q}{wh^2}, \tag{1}$$

where $\dot{\gamma}_{app}$ is the apparent shear rate of flow at the wall, and $w$ and $h$ are the width and the height of the channel respectively (2 mm × 50 μm). The actual shear rate at the wall, $\dot{\gamma}$, is related to $\dot{\gamma}_{app}$ as[25]

$$\dot{\gamma} = \frac{\dot{\gamma}_{app}}{3}\left[2 + \frac{d\ln(\dot{\gamma}_{app})}{d\ln(\tau)}\right] = \dot{\gamma}_{app}\left[\frac{2}{3} + \frac{1}{3}\frac{d\ln(Q)}{d\ln(\Delta p)}\right]. \tag{2}$$

For Newtonian fluids and laminar flow, $\tau$, shear stress, is linearly proportional to $\dot{\gamma}_{app}$, so $\dot{\gamma} = \dot{\gamma}_{app}$. For non-Newtonian fluids, the shear rate needs to be modified using Eq. (2) to calculate the true viscosity. Four pressure sensors are mounted at the boundary wall in the channel to detect the pressure drop from the inlet to the outlet as a function of position along the channel, and the shear stress, $\tau$, is calculated from the pressure drop as[25]

$$\tau = \frac{\Delta p}{L} \frac{wh}{2w+2h}, \qquad (3)$$

where $\Delta p$ is the measured pressure drop, and $L$ is the channel length over which $\Delta p$ is measured (15 mm). Two different microchips were used, each with the same dimensions but different pressure measurement ranges, having maximum measurable pressure drops of $\Delta p_{max}$ = 10,000 Pa and $\Delta p_{max}$ = 40,000 Pa respectively, and with a minimum measurable pressure drop of 1% of the maximum. These limits bound the range of measurable viscosity versus shear rate. The dynamic viscosity is defined as[42]

$$\eta \equiv \frac{\tau}{\dot{\gamma}}. \qquad (4)$$

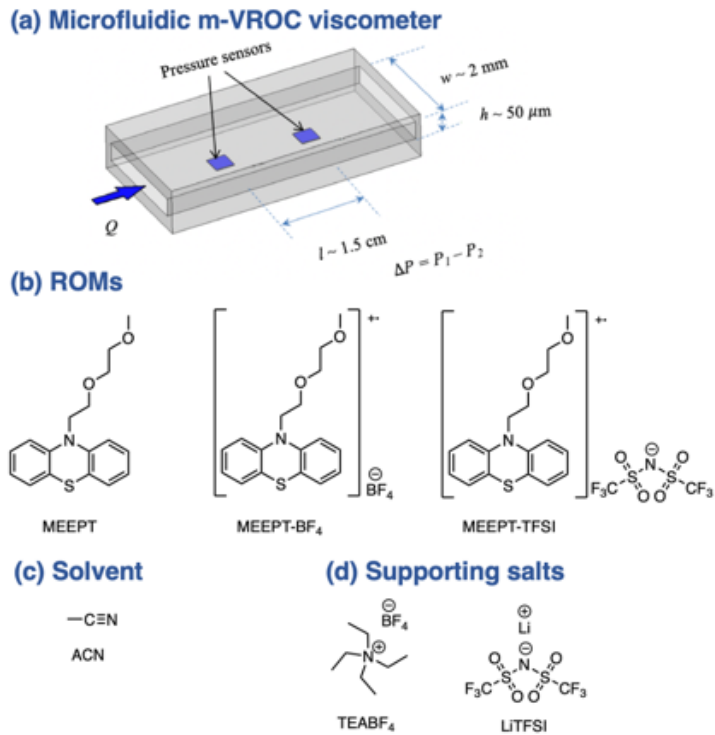

**Figure 1.** Representation of (a) microfluidic viscometer (image adapted from RheoSense Inc.), (b) ROMs: MEEPT (neutral), MEEPT-BF$_4$ (charged), and MEEPT-TFSI (charged), (c) solvent: acetonitrile (ACN), and (d) supporting salts: TEABF$_4$ and LiTFSI.

Experimental limits constrain the range of $\dot{\gamma}$ available to explore non-Newtonian behavior because of the maximum and minimum pressure drop that the sensors can measure, and the maximum flow rate. Such limits are converted to experimental windows,[40] as shown in the viscosity versus shear rate plots. For each solution, viscosities were measured within a wide shear rate range (5,000-30,000 s$^{-1}$) for most samples (varied depending on the viscosity of each solution). The temperature was maintained at 25 ºC by a Thermocube circulator. For each sample, repeat measurements were done (typically in triplicate), and the uncertainly was calculated as the standard deviation.

**Table 1.** The solubilities and composition of ROMs in different supporting salt environments investigated in this study.

| ROM | Electrolyte | Solubility (M) | Maximum concentrations tested in this study (M) |
|---|---|---|---|
| MEEPT | ACN | miscible | 3.78 |
|  | 0.5 M TEABF$_4$/ACN | miscible | 1.00 |
|  | 0.5 M LiTFSI/ACN | miscible | 1.00 |
| MEEPT-BF$_4$ | ACN | 0.55 | 0.45 |
|  | 0.5 M TEABF$_4$/ACN | 0.45 | 0.36 |
|  | 0.5 M LiTFSI/ACN | 0.81 | 0.43 |
| MEEPT-TFSI | ACN | 1.54 | 1.00 |
|  | 0.5 M TEABF$_4$/ACN | 1.11 | 1.00 |
|  | 0.5 M LiTFSI/ACN | 1.18 | 1.00 |

## C. Viscosity models

To describe the concentration-dependent viscosity, a few equations are established here. For solutions containing uncharged hard spheres, Einstein formulated an equation to theoretically calculate the viscosity of dilute molecular solutions of rigid, non-attracting spheres at different concentrations,[33] as

$$\frac{\eta(\phi)}{\eta_s} = 1 + \frac{5}{2}\phi + O(\phi^2), \qquad (5)$$

where $\eta(\phi)$ and $\eta_s$ are the viscosities of solution and solvent respectively, and $\phi$ is the volume fraction of solute particles. For spherical particles, $\phi$ can be written as

$$\phi = \frac{4}{3}\pi\left(\frac{d_H}{2}\right)^3 cN_A, \qquad (6)$$

where $d_H$ is the hydrodynamic diameter of an individual particle, $c$ is the concentration (moles per volume), and $N_A$ is Avogadro's constant. Thus, the Einstein equation can be written in terms of concentration, as

$$\frac{\eta(c)}{\eta_s} = 1 + [\eta]c + O(c^2), \qquad (7)$$

where $[\eta]$ is the intrinsic viscosity, which is a function of only hydrodyanmic diameter, $d_H$, such that

$$[\eta] = \frac{10}{3}\pi\left(\frac{d_H}{2}\right)^3 N_A. \qquad (8)$$

Experimentally, Huggins expanded this equation to a more concentrated regime, with a higher-order term representing the particle interactions,[34]

$$\frac{\eta(c)}{\eta_s} = 1 + [\eta]c + k_H ([\eta]c)^2 + O(c^3) \qquad (9)$$

which is called the Huggins equation, where $k_H$ is the Huggins coefficient that quantifies the interactions between particles. For Brownian hard spheres with only hydrodynamic interactions and steric effects, Batchelor numerically derived the second-order coefficient[36] to be

$$\frac{\eta(\phi)}{\eta_s} = 1 + \frac{5}{2}\phi + 6.2\phi^2, \qquad (10)$$

which corresponds to $k_H = 0.992$.

For salt solutions, the viscosity is most commonly described by the Jones-Dole equation,[35] which is

$$\frac{\eta(c)}{\eta_s} = 1 + A\sqrt{c} + Bc. \qquad (11)$$

This can be extended to higher concentrations as

$$\frac{\eta(c)}{\eta_s} = 1 + A\sqrt{c} + Bc + Dc^2, \qquad (12)$$

where $A$, $B$, $D$ are fit coefficients and $A\sqrt{c}$ represents the interactions and mobility of ions, which can be calculated using the Falkenhagen theory.[43,44] Here the term $Bc$ represents ion-solvent interactions, and $Dc^2$ represents ion-dipole interactions and long-range coulombic ion-ion interactions.

In this study, the Einstein, Huggins, and extended Jones-Dole equations were used to describe the measured viscosities of neutral MEEPT and two radical cation salts in different supporting salt environments. These equations are preferred over others, because they are the most credible models for each solution respectively in the concentrations studied. The credibility was determined from the Bayesian Information Criterion (BIC), which is introduced in the next section.

### D. Fit method and credibility

The fit and model selection were done using the "fitnlm" function in MATLAB,[45] a commonly employed nonlinear regression model that uses an iterative generalized least squares algorithm, which in its most generalized form minimizes the residual sum of squares (RSS).[46] The weighting function was specified as the experimental uncertainty (standard deviation) of each measured data point.

To evaluate the hydrodynamic diameter and quantify the interaction of molecules in solution, the intrinsic viscosity and Huggins coefficient can be fit to the measured viscosities at given concentrations. However, they are defined in the limit of $c \to 0$,[47] while all the measurements are done at a finite concentration. This is a known issue and various approaches have been used to solve this conflict, such as single point method and extrapolation.[48] Here, the extrapolation method is conducted by rearranging the Huggins equation as

$$\eta_{red} = \frac{\eta/\eta_s - 1}{c} = [\eta] + k_H [\eta]^2 c, \qquad (13)$$

where $\eta_{red}$ is called the reduced viscosity. The reduced viscosity versus concentration is fit to a linear line and the intrinsic viscosity is evaluated as the intercept at zero concentration. Since the viscosity contributed by the solvent must be subtracted before extrapolation, small experimental errors can be amplified,[49] and this is addressed by including a weighting function in the fitting based on propagated uncertainty of the data. The most credible fit is found by choosing the appropriate number of data points used for fit, and the selection is done by comparing the Baysesian Information Criterion (BIC) of each fit result.

BIC is a criterion for model selection, which takes both the goodness of fit and the penalty term for the number of parameters in the model into account. It is an approximation of the full Bayes factors used to assess model credibility. While the full Bayesian calculation has been applied previously to complex fluid rheology,[50] the BIC approximation is much less computationally involved. For each fit it is calculated as

$$\mathrm{BIC} = -2\ln L + k\ln n, \tag{14}$$

where $\ln L$ is the log-likelihood, which represents the goodness of the fit, $n$ is the number of data points used for fit, and $k$ is the number of parameters fit. The second term on the right side is the penalty term acounting for over-parameterizing. The fit with the lowest BIC is preferred.

The viscosities of MEEPT solutions were measured and analyzed using the methods mentioned above, and the results are discussed next.

## 3. RESULTS AND DISCUSSION

### A. MEEPT: a flowable ROM

A wide range of concentrations of MEEPT/ACN solutions and neat MEEPT were studied. Neat MEEPT is a flowable, yellow, oily liquid, as shown in the photograph inset in Figure 2(a). It is miscible with ACN in any proportion. The viscosities of neat MEEPT and MEEPT/ACN solutions were measured over a wide range of shear rates, and Newtonian viscosities were observed for all samples within 5% variation, as shown in Figure 2(a). The error bars come from the standard deviation for repeat measurements.

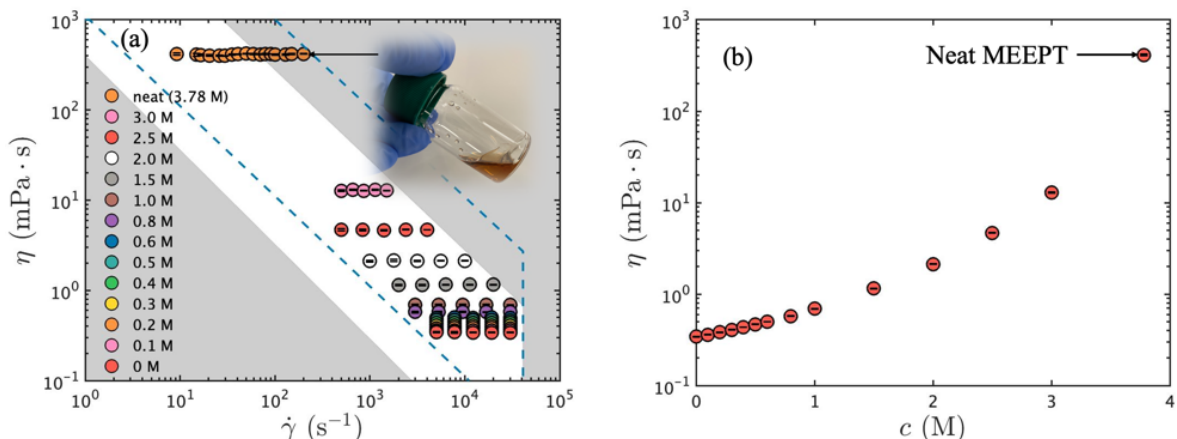

**Figure 2.** Viscosities of neat liquid MEEPT and MEEPT/ACN solutions at different concentrations (0 – 3 M): (a) the dynamic viscosities as a function of shear rate $\dot{\gamma}$; the gray areas are the experimental limits for the measuring chip used for MEEPT/ACN solutions, and the blue dashed lines are drawn to show the limits for the chip used for neat MEEPT, which are due to the maximum and minimum pressure drops that the pressure sensors can measure in the flow channel, (b) the average Newtonian viscosity as a function of MEEPT concentration. Note: the concentration of neat MEEPT is around 3.8 M.

The average Newtonian viscosity at each concentration was calculated by taking an average over viscosities at different shear rates; the error bars come from both the standard deviation and the uncertainty propagation, as shown in Figure 2(b). The viscosity of neat MEEPT is 412 mPa·s, 1,000 times larger than that of ACN. Therefore, as the MEEPT concentration increases, the viscosity of MEEPT/ACN solutions increases, first linearly in the dilute regime, then more dramatically when the concentration becomes larger than 1 M. Note that when the concentration is 1 M, the viscosity is around 0.7 mPa·s, twice of that of the solvent itself, which is not a large increase and would not have a destructive impact on the ionic conductivity, suggesting that the solution remains "flowable" up to 1 M.

## B. MEEPT in different supporting salt environments

In RFB electrolytes, the ionic conductivity of the nonaqueous solvent, ACN in this case, is too low to support effective charging, so supporting salts are added to raise ionic conductivity and to provide counterions for charged ROMs. For MEEPT in ACN electrolytes containing supporting salts, the conformation and interactions between MEEPT might change, which affects the viscous flow properties of MEEPT solutions. Thus, viscosities of MEEPT in different supporting salt envrionments are studied.

In this paper, two supporting salts are used: TEABF$_4$ and LiTFSI. Their concentration-dependent viscometric analysis is shown in the SI. In Figure S2, at 0.5 M, the viscosities of TEABF$_4$/ACN and LiTFSI/ACN solutions are 0.45 and 0.50 mPa·s, respectively. The 0.5 M TEABF$_4$ and 0.5 M LiTFSI were used separately as supporting salts in electrolytes to test the viscous flow properties of MEEPT. The measurement results, which show Newtonian behavior within 3% variation, are shown in the Figure 3.

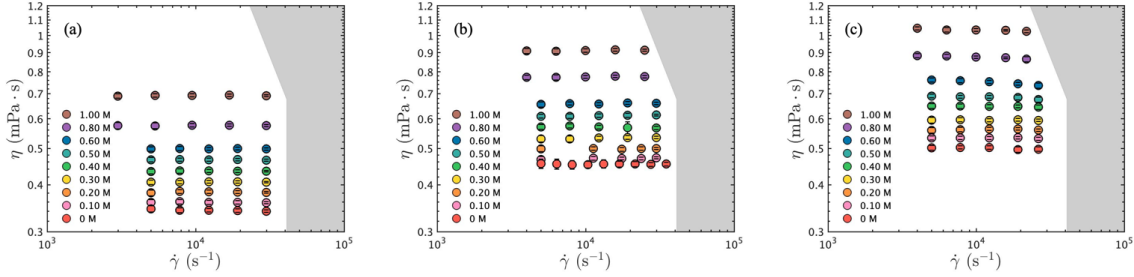

**Figure 3.** The dynamic viscosities of MEEPT/ACN solutions in different supporting salt environments, (a) no salt, (b) TEABF$_4$ (0.5 M), (c) LiTFSI (0.5 M); the viscosities show Newtonian behavior within 3% variation.

Similarly, the concentration-dependent average Newtonian viscosities for both solutions were calculated and plotted in comparison with MEEPT/ACN solutions, as shown in Figure 4(a). It can be seen that at zero MEEPT concentration, the viscosities of the three electrolytes are different, with 0.5 M LiTFSI/ACN > 0.5 M TEABF$_4$/ACN > ACN without supporting salt, but the increasing slopes are similar. To quantify the effect of supporting salts on the increase of viscosity versus MEEPT concentration, the reduced viscosity is plotted as a function of concentration (0.1 – 1.0 M), as shown by the data points in Figure 4(b), and is fit with the rearranged Huggins equation (Eq. (13)).

In the Huggins equation (Eq. (9)), only two factors are taken into account, namely the excluded volume described by the linear term, $[\eta]c$, and the interactions between particles, represented by the square term, $k_H ([\eta]c)^2$. If higher order terms can be neglected, the reduced viscosity, $\eta_{red}$, as shown in the rearranged Huggins equation (Eq. (13)), is a linear function of $c$. First, to determine the critical concentration at which the higher order terms become important and the Huggins equation no longer holds, the most credible concentration range for the Huggins equation fit was found. This was done by fitting the reduced viscosity versus concentration data to the rearranged Huggins equation (Eq. (13)) with different numbers of data points. By changing the maximum concentration used to fit, that is, 3 points from 0.1 to 0.3 M were fit to the equation, then 4 points from 0.1 to 0.4 M, and so on, until the maximum concentration tested was used. The BIC of each fit were compared, and the one with minimum BIC is judged to be the most credible fit, and the coresponding concentration range is the most appropriate range for the Huggins equation, beyond which the equation is no longer suitable. The fit results at each sample size are shown in the SI, and the fit lines over different concentration ranges are ploted in Figure 5. From fit results for MEEPT/ACN with no salt (red), the deviation of fit results using data at sufficiently high concentrations can be seen, because at that high concentration, the viscosity increase is more dramatic and the Huggins equation no long holds. From fit for MEEPT/ACN solutions with 0.5 M TEABF$_4$ (orange) and with 0.5 M LiTFSI (green), the fit lines using fewer data points at low concentration are oberserved to deviate from the data, due to the large uncertainty of the measurements at low concentration regimes. The BIC of each fit is plotted versus the number of data points used for fitting. It can be seen that for MEEPT/ACN with no salt (red), the BIC decreases with the number of data points, reaching its minimum at $n = 7$, which corresponds to concentrations ranging from 0.1 to 0.8 M, then increases as the number becomes larger. This means that until 0.8 M, the higher order term in the Huggins equation can be neglected and it is appropriate to use it to describe the concentration-dependent viscosity. For solutions with 0.5 M TEABF$_4$ (orange) and with 0.5 M LiTFSI (green), the BIC decrerases with $n$, reaching its minimum at $n=8$, which is 0.1 – 1.0 M. The most credible fit results are shown in the solid line in Figure 4, where in the reduced viscosity plot, the intercept represents the intrinsic viscosity, $[\eta]$, and the Huggins coefficient, $k_H$, is determined from the slope. Therefore, the hydrodynamic diameters, $d_H$, of MEEPT can be determined from $[\eta]$ using Eq. (8), which are $8.52\pm0.07$, $7.72\pm0.12$, and $8.56\pm0.16$ Å for MEEPT/ACN solutions without supporting salts, with 0.5 M 0.5 M TEABF$_4$, and with 0.5 M LiTFSI, respectively. Table 2 summarizes these results.

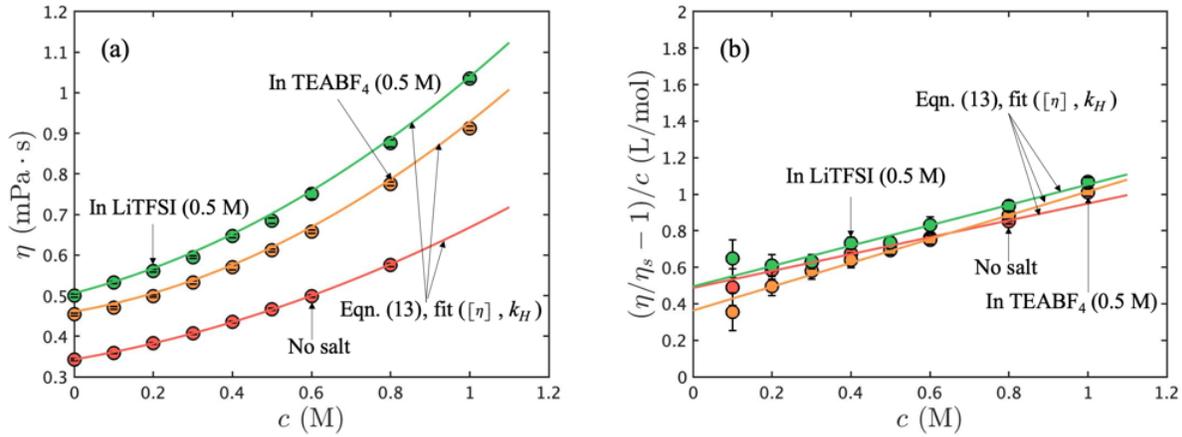

**Figure 4.** The viscometric analysis results of MEEPT/ACN solutions with no salt (red), 0.5 M TEABF$_4$ (orange), and 0.5 M LiTFSI (green) as a function of MEEPT concentration. The solid lines show the rearranged Huggins equation (Eq. (13)) fit results; (a) the average Newtonian viscosities, (b) the reduced viscosities; intercept is the intrinsic viscosity, $[\eta]$, and slope is proportional to the Huggins coefficient, $k_H$.

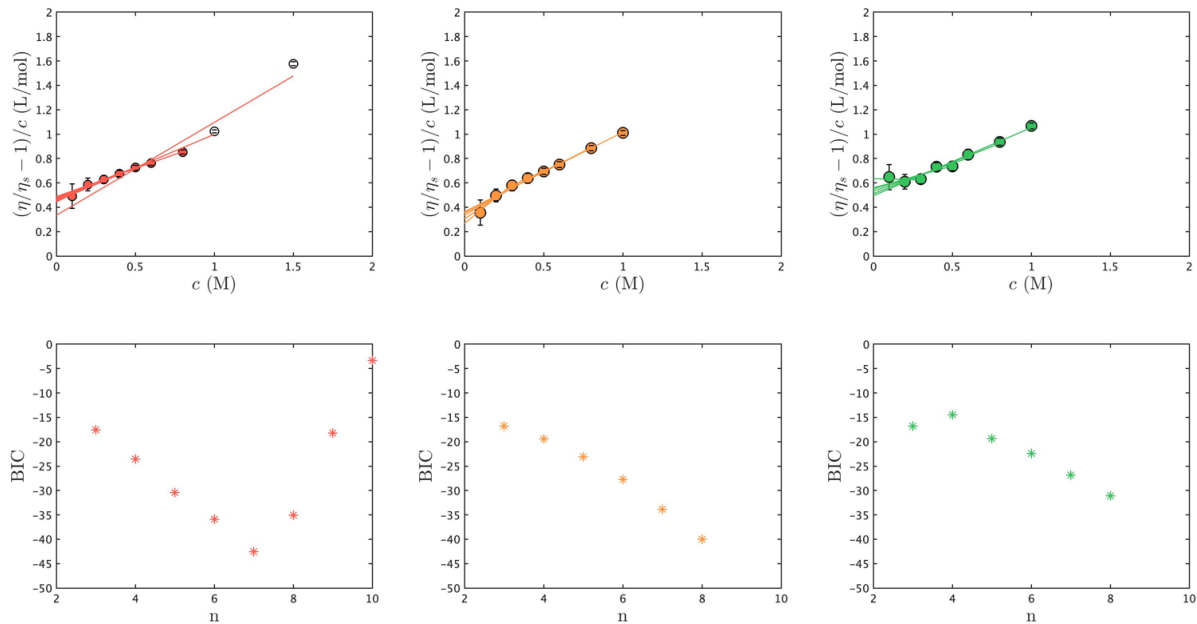

**Figure 5.** Fit procedure and selection of MEEPT/ACN solutions with no salt (red), 0.5 M TEABF$_4$ (orange) and 0.5 M LiTFSI (green): the reduced viscosities as a function of MEEPT concentrations, all the data are used to fit the rearranged Huggins equation (Eq. (13)), from which the hydrodynamic diameters and Huggins coefficients are determined. The credibility of fit: BIC as a function of number of data points ($n$) fit to the Eq. (13). For MEEPT/ACN solutions with no salt (red), $n = 7$ is the most credible fit according to BIC, for solutions with 0.5 M TEABF$_4$ (orange) and with 0.5 M LiTFSI (green), $n = 8$ are the most credible fits.

**Table 2.** MEEPT in different supporting salt environments: the most credible fit results of concentration-dependent reduced viscosity to the rearranged Huggins equation (Eq. (13)).

|  | $c_{max}$ (M) | $\eta_s$ (mPa·s) | $[\eta]$ (L/mol) | $d_H$ (Å) | $k_H$ (-) |
|---|---|---|---|---|---|
| No salt | 0.80 | 0.34 ± 0.01 | 0.49 ± 0.01 | 8.52 ± 0.07 | 1.95 ± 0.10 |
| TEABF$_4$ (0.5 M) | 1.00 | 0.46 ± 0.01 | 0.36 ± 0.02 | 7.72 ± 0.12 | 4.94 ± 0.29 |
| LiTFSI (0.5 M) | 1.00 | 0.50 ± 0.01 | 0.49 ± 0.03 | 8.56 ± 0.16 | 2.28 ± 0.21 |

For comparison of the hydrodynamic diameters determined from viscometric analysis, four representative geometries of the neutral MEEPT molecule were optimized using Density Functional Theory (DFT) with the B3LYP/6-311G** basis set, as implemented in Gaussian 16.[51] All optimized structures were verified by frequency calculations to be local minima. To determine the lowest energy conformations of the MEEPT molecule, we sampled four initial configurations with different rotating angles of the 2-(2-methoxyethoxy)ethyl group relative to the C atoms *para* to the N atoms in the phenothiazine ring system. The optimized structures and their relative energies are shown in Figure 6. The energy differences of the four optimized MEEPT molecules are within 0.2 eV. The dimensions of MEEPT molecules are specified in Figure 6.

Compared with the molecular structure dimensions, which is about 7 to 11 Å, the hydrodynamic diameters determined from viscosity measurements are comparable, and are almost equal within the experimental error in different supporting salts. This means that the addition of supporting salts does not have a significant effect on the hydrodynamic diameter of MEEPT, and there is minimal flocculation in these solutions under the conditions tested.

The Huggins coefficients, which quantify the molecular interactions, are 1.95±0.10, 4.94±0.29, and 2.28±0.21, respectively, from the most credible fits to the rearranged Huggins equation. Compared with 0.992, which is the Huggins coefficient derived by Batchelor for Browian hard spheres with only hydrodynamic inteactions and steric effects, those for MEEPT in different supporting salts are larger. That suggests that there are extra interactions between MEEPT that contribute to the viscous dissipation at finite concentrations.

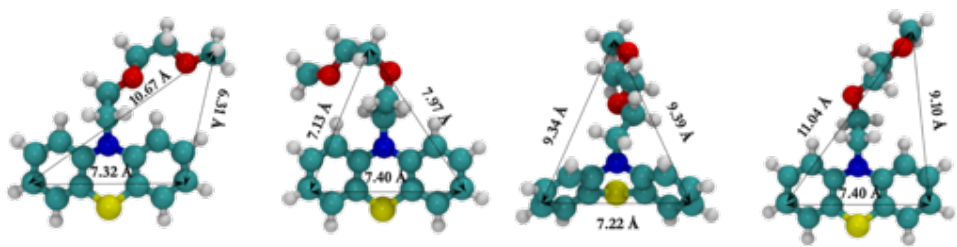

**Figure 6.** Four optimized geometries and relative dimensions of neutral MEEPT. Blue, yellow, cyan, red, and white spheres denote the nitrogen, sulfur, carbon, oxygen, and hydrogen atoms, respectively. The relative energies of the four geometries from left to right are 0, 0.04, 0.05, and 0.16 eV, respectively.

MEEPT solutions containing different supporting salts represent the uncharged electrolytes in RFBs; when charged, solution viscosities are unknown because of the excluded volume of anion and the effects of solubility limits of its charged form have not been quantified so far. The impact of supporting salts on the viscous flow properties of the MEEPT radical cation salt solutions, which represent the charged solutions, is also unknown. The viscous flow properties of MEEPT radical cation salt solutions are addressed next.

### C. Charged MEEPT cations

The two radical cation salts of MEEPT are investigated here: MEEPT-BF$_4$ and MEEPT-TFSI. The species have solubilities around 0.5 M and 1 M, respectively, varying within 60% in different supporting salt environments, as shown in Table 1. MEEPT-TFSI is more soluble than MEEPT-BF$_4$ in all three conditions. For each radical cation, the addition of supporting salt changes the solubility, and different supporting salts have different effects; TEABF$_4$ reduces solubility more dramatically for species.

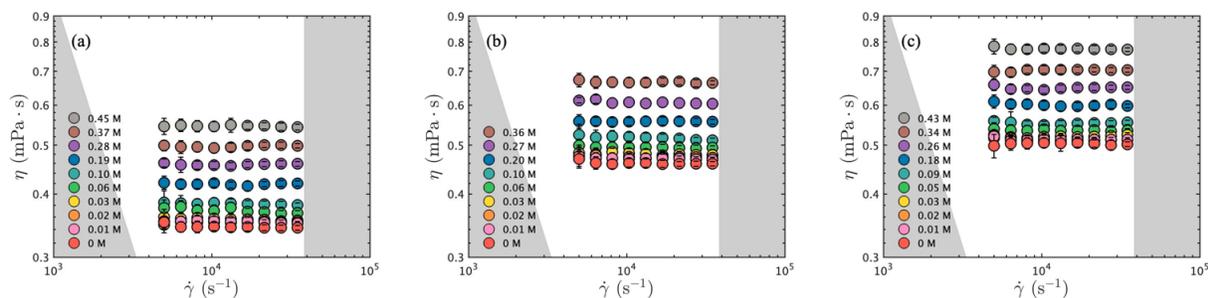

**Figure 7.** The dynamic viscosities of MEEPT-BF$_4$/ACN solutions in different supporting salt environments (a) no salt, (b) TEABF$_4$ (0.5 M), and (c) LiTFSI (0.5 M); the solutions exhibit Newtonian behavior within 3% variation.

Results from viscosity measurements for MEEPT-BF$_4$ in three different environments (ACN, 0.5 M TEABF$_4$/ACN, and 0.5 M LiTFSI/ACN), are shown in Figure 7. The solutions are Newtonian within 3% variations. The average Newtonian viscosities at different concentrations were calculated and plotted in Figure 8, with the error bars coming from standard deviation and uncertainty propagation. Even though at zero MEEPT-BF$_4$ concentration, the viscosities of the three solutions are different in the presence of supporting salts. The slopes of MEEPT-BF$_4$ viscosities are almost the same for all three solutions. For MEEPT-BF$_4$ concentrations up to 0.5 M, the viscosities increase less than a factor of two, and thus the solutions remain "flowable".

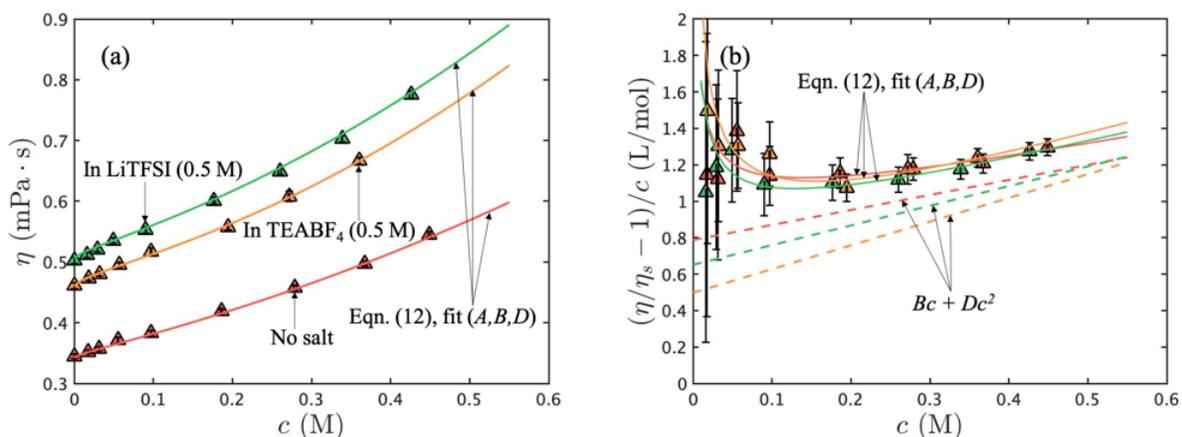

**Figure 8.** The viscometric analysis results of MEEPT-BF$_4$/ACN solutions with no salt (red), 0.5 M TEABF$_4$ (orange), and 0.5 M LiTFSI (green) as a function of MEEPT-BF$_4$ concentration: (a) the average Newtonian viscosity, (b) the reduced viscosity, the solid lines are fit to the rearranged Huggins equation, and the dash lines are the fit results without $A\sqrt{c}$ term, where the intercept is the intrinsic viscosity, $[\eta]$, and the slope is proportional to the Huggins coefficient, $k_H$.

For charged cation solutions, the measured viscosities are fit to the Einstein (Eq. (5)), the Huggins (Eq. (9)), the Jones-Dole (Eq. (11)), and the extended Jones-Dole equations (Eq. (12)). The details are shown in the SI, and from the Table S3, the extended Jones-Dole equation fits have the lowest BIC for all three solutions, and the fit results are shown as solid lines in Figure 8.

**Table 3.** MEEPT-BF$_4$ in different supporting environments, the most credible fit results of concentration-dependent average Newtonian viscosity to the extended Jones-Dole equation (Eq.(12)).

| | $c_{max}$ (M) | A (L$^{1/2}$/mol$^{1/2}$) | B (L/mol) | D (L$^2$/mol$^2$) | $d_H$ (Å) | $k_H$ (-) |
|---|---|---|---|---|---|---|
| No salt | 0.45 | 0.08 ± 0.04 | 0.79 ± 0.14 | 0.82 ± 0.20 | 10.00 ± 0.59 | 1.33 ± 0.40 |
| LiTFSI (0.5 M) | 0.42 | 0.10 ± 0.03 | 0.65 ± 0.10 | 1.08 ± 0.17 | 9.38 ± 0.52 | 2.54 ± 0.57 |
| TEABF$_4$ (0.5 M) | 0.36 | 0.16 ± 0.05 | 0.50 ± 0.20 | 1.30 ± 0.36 | 8.58 ± 1.15 | 5.27 ± 2.56 |

From Figure 8(b), it can be seen that unlike MEEPT, the reduced viscosities of MEEPT-BF$_4$ decrease at first, then increase linearly, as if the apparent hydrodynamic diameters of MEEPT-BF$_4$ decrease at first, which also suggests that a square root term should be included in the viscosity model. In the Jones-Dole equation, the $A\sqrt{c}$ term represents the interaction and mobility of ions and this term becomes less influential as concentration increases. The dashed lines in Figure 8(b) are the extended Jones-Dole equation fit results without the $A\sqrt{c}$ term, and it can be seen that the two lines differ in the low concentration regime, then converge as the concentration increases. Thus, the $B$ coefficient is interpreted as an intrinsic viscosity, and the hydrodynamic diameter and the Huggins coefficient of MEEPT-BF$_4$ are evaluated from the $B$ and $D$ coefficients, as shown in Table 3. The hydrodynamic diameters of MEEPT-BF$_4$ are 10.00 ± 0.59, 9.38 ± 0.52, and 8.58 ± 1.15 Å, respectively in the three electrolytes, with Huggins coefficients of 1.33 ± 0.40, 2.54 ± 0.5, and 5.27 ± 2.56 respectively. Therefore, we can conclude that the addition of supporting salts decreases the hydrodynamic diameter of MEEPT-BF$_4$, while promoting the interactions between MEEPT-BF$_4$ species. These results are summarized in Table 3.

In addition to MEEPT-BF$_4$, we also analyzed MEEPT-TFSI, which is more soluble than MEEPT-BF$_4$, as shown in Table 1. The results of its viscosity measurements are shown in Figure 9 in the three electrolyte environments: ACN, 0.5 M TEABF$_4$/ACN, and 0.5 M LiTFSI/ACN. The viscosities of MEEPT-TFSI in different solutions show Newtonian behavior even at 1 M, within 4% variation. Therefore, average Newtonian viscosities were calculated and plotted for the concentration-dependence analysis.

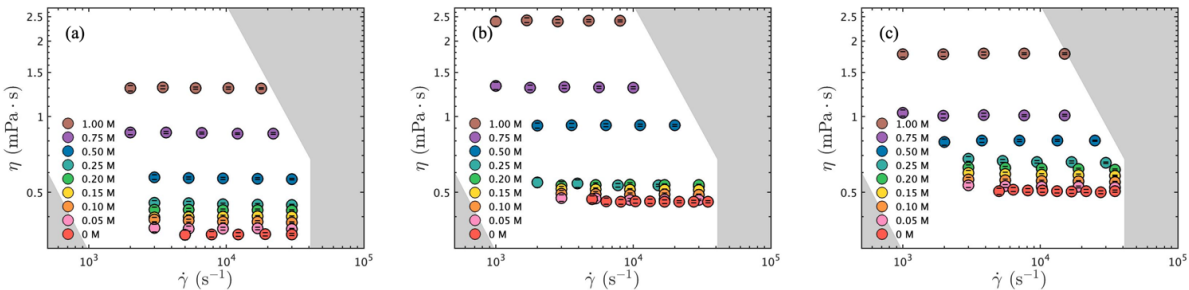

**Figure 9.** The dynamic viscosities of MEEPT-TFSI/ACN solutions in different supporting salt environments: (a) no salt, (b) TEABF$_4$ (0.5 M), and (c) LiTFSI (0.5 M); the solutions exhibit Newtonian behavior within 4% variation.

Figure 10 shows the average Newtonian viscosities and reduced viscosities of MEEPT-TFSI in the three electrolytes. Similar to MEEPT-BF$_4$, at concentrations up to 0.5 M, the viscosities of radical cation solutions increase slowly, and the solutions remain "flowable". However, as the concentration further increases, the higher order effects become significant and the viscosities increase dramatically, as MEEPT-TFSI approaches its maximum solubility limits.

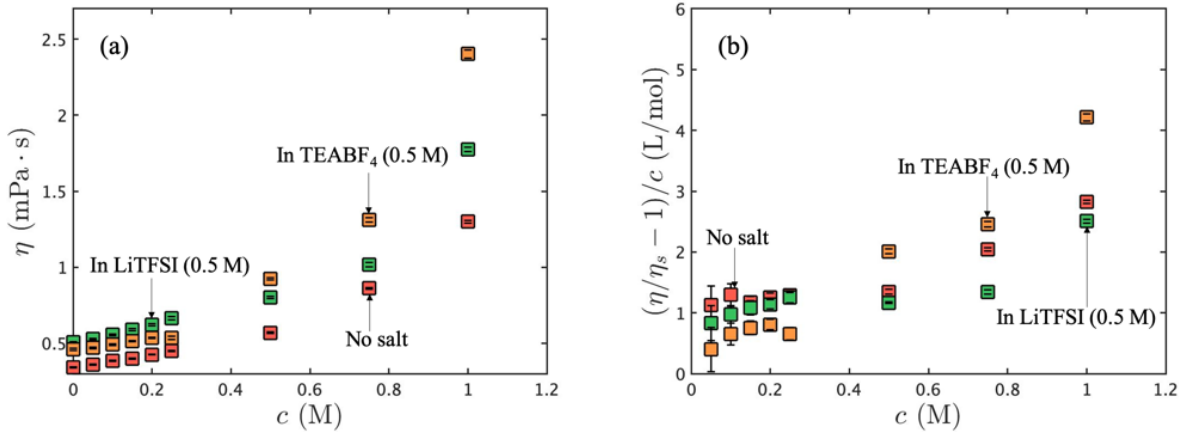

**Figure 10.** The viscosities of MEEPT-TFSI/ACN solutions with no salt (red), with 0.5 M TEABF$_4$ (orange), and with 0.5 M LiTFSI (green), as a function of MEEPT-TFSI concentration: (a) the average Newtonian viscosities, (b) the reduced viscosities.

From Figure 10(b), we can see that the reduced viscosities show increasing trend in the low concentration regime, which suggests an increasing apparent hydrodynamic diameter and a negative $A$ coefficient in the Jones-Dole equation, which contradicts Falkenhagen theory.[43,44] Currently, we are unable to describe such concentration-dependent viscosities using any equation available. Therefore, the apparent hydrodynamic diameter, $d_{app} = \left( \frac{12B}{5\pi N_A} \right)^{1/3}$, is evaluated at a single concentration of 0.2 M, where the $A\sqrt{c}$ terms become insignificant and the reduced viscosities begin to increase linearly with concentration. The results are shown in Table 4.

**Table 4.** MEEPT-TFSI in different supporting environments: the apparent hydrodynamic diameters of MEEPT-TFSI at 0.20 M.

|  | $d_{app}$* (Å) |
| --- | --- |
| No salt | 11.68 ± 0.22 |
| TEABF$_4$ (0.5 M) | 10.08 ± 0.37 |
| LiTFSI (0.5 M) | 11.31 ± 0.26 |

* $d_{app}$ means the apparent hydrodynamic diameter at 0.2 M.

From the hydrodynamic diameter calculations, we saw that the diameters for both MEEPT-BF$_4$ and MEEPT-TFSI increase compared to the neutral MEEPT and remain almost unchanged with addition of supporting salts. The increase of hydrodynamic diameters might be due to the excluded volume of anions, which can be seen in Figure 11. The DFT-optimized models show that the molecular structure dimensions of MEEPT-BF$_4$ and MEEPT-TFSI are a little larger compared with neutral MEEPT. The two ion-pair geometries were optimized using the same method as with neutral MEEPT. The starting geometries of the MEEPT radical cation salt species were taken from the their thermal ellipsoid plots of single crystal X-ray structures.[32,37] These initial configurations were also justified by examining the positive charge distribution on the MEEPT radical cation: the positive charge is located at the fragment of the molecule facing the negative counter-ion and thus these configurations are preferential.

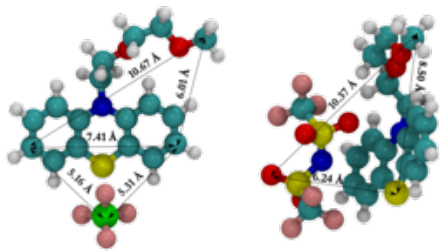

**Figure 11.** Optimized geometries of MEEPT-BF$_4$ and MEEPT-TFSI ion pair and their dimensions. Blue, yellow, cyan, red, white, pink, and green balls denote the nitrogen, sulfur, carbon, oxygen, hydrogen, fluorine, and boron atoms.

When the concentration of MEEPT-TFSI becomes larger than 0.5 M. the viscosities of solutions increase dramatically, which means the interactions between MEEPT-TFSI and MEEPT-TFSI with supporting salts become more significant. In our work with other ROMs and supporting salt electrolytes, we have shown both experimentally and theoretically that the concentration of maximum ionic conductivity is associated with the concentration at which the viscosity begins to increase non-linearly,[52] which is about 0.5 M in this case. Even though the viscosity of 0.5 M TEABF$_4$/ACN is smaller than that of 0.5 M LiTFSI/ACN, it can be seen from Figure 10(a) that when MEEPT-TFSI concentrations exceed 0.5 M, its viscosity in 0.5 M TEABF$_4$ is larger than that in 0.5 M LiTFSI, suggesting the interactions in 0.5 M TEABF$_4$/ACN are stronger compared to in 0.5 M LiTFSI/ACN.

For MEEPT/ACN solutions (Figure 4), we observed that the viscosity increases linearly with concentration up to 0.5 M, no matter what supporting salts are used, and that viscosity increases by less than a factor of two, up to 1.0 M. However, for MEEPT-TFSI/ACN solutions (Figure 10), the viscosity increases dramatically at concentrations exceeding 0.5 M, which means that in the operation of NAqRFBs using MEEPT as the ROM, when MEEPT solutions are charged, the reconstitution of fluids at such high concentrations can greatly affect the viscous flow properties of the electrolyte, including viscosity, diffusion, and ionic conductivity.

Table 5 summarizes the results for all solutions tested.

**Table 5.** Summary of viscosity and hydrodynamic diameter results.

| ROM | Electrolyte | Hydrodynamic diameter (Å) | Viscosity (mPa·s) | Concentration (M) | Analyzed by |
|---|---|---|---|---|---|
| MEEPT | ACN | 8.52 ± 0.07 | 0.58 ± 0.01 | 0.80 | Huggins equation |
|  | 0.5 M TEABF$_4$/ACN | 7.72 ± 0.12 | 0.91 ± 0.01 | 1.00 |  |
|  | 0.5 M LiTFSI/ACN | 8.56 ± 0.16 | 1.03 ± 0.01 | 1.00 |  |
| MEEPT-BF$_4$ | ACN | 10.00 ± 0.59 | 0.45 ± 0.01 | 0.45 | Extended Jones-Dole equation |
|  | 0.5 M TEABF$_4$/ACN | 9.38 ± 0.52 | 0.67 ± 0.01 | 0.42 |  |
|  | 0.5 M LiTFSI/ACN | 8.58 ± 1.15 | 0.78 ± 0.01 | 0.36 |  |
| MEEPT-TFSI | ACN | 11.68 ± 0.22 | 0.43 ± 0.01 | 0.20 | Einstein equation one-point method |
|  | 0.5 M TEABF$_4$/ACN | 11.31 ± 0.26 | 0.54 ± 0.01 | 0.20 |  |
|  | 0.5 M LiTFSI/ACN | 10.08 ± 0.37 | 0.62 ± 0.01 | 0.20 |  |

## D. Discussion to rationalize Newtonian behavior

With the inference of hydrodynamic diameter in all the tested compositions, we can *a posteriori* rationalize the observation of Newtonian behavior up to shear rates of 30,000 s$^{-1}$. We do this by considering an estimate of a diffusion-based relaxation time[53,54] for a single dilute MEEPT molecule,

$$\tau = 6\pi \eta_s r_H^3 / (k_B T) \tag{15}$$

where $k_B T$ is the thermal energy, and postulating that non-Newtonian effects in steady shear become significant only when the Weissenberg number, defined as

$$Wi = \tau \dot{\gamma} \tag{16}$$

becomes order unity. This number expresses the ratio between the rate at which the structure of the particle distribution is deformed by shear flow and the rate of Brownian diffusion that helps recover the equilibrium conformation. Weissenberg number is a general concept for fluid nonlinearity,[55] and here it relates specifically to the concept of a Peclet number, a ratio of advection rate to diffusion rate. We note that the expression in Eq. (15) the time scale $\tau$ is a lower-bound estimate, since $\tau$ would increase sensitively with size (e.g. association of multiple molecules), increase linearly with background viscosity, and increase significantly at higher concentrations when intermolecular interactions become important.

For ACN solutions, of which $\eta_s = 0.34$ mPa·s, containing MEEPT with known hydrodynamic dimeter $d_H = 8.5$ Å, with negligible Coulomb or van der Waals forces, flowing at a shear rate, $\dot{\gamma}$, ranging from 5,000 to 30,000 s$^{-1}$, and temperature, $T$ = 298 K, the corresponding Weissenberg (Peclet) number ranges from $6\times10^{-7}$ to $4\times10^{-6}$, much less than unity. From this, the solutions are expected to be Newtonian, as observed in our experiments. To show shear-thinning behavior, the Weissenberg (Peclet) number must be unity or larger. We can define a critical shear rate for the condition Wi = 1, yielding for our specific case the critical shear rate

$$\dot{\gamma}_c = 8.4\times10^9 \text{ s}^{-1}, \tag{17}$$

which is much larger than the experiment limits. Thus, in our study no shear-thinning is observed for any solution. Larger molecules, or aggregates of molecules, would change this estimate. For example, for a critical shear rate of 10,000 s$^{-1}$ (within range of the instrument used here), the critical diameter is around 80 nm, which is about one hundred times larger than a single MEEPT molecule. From this analysis, the lack of shear-thinning implies that no structures of such size are contained within the complex fluid compositions tested here.

## 4. CONCLUSIONS

In summary, the viscous flow properties of MEEPT and its radical cation form were studied in different supporting salt environments, and it is shown that solution viscosity of both neutral and charged MEEPT increase as a result of excluded volume and species interactions; the excluded volume is comparable to the molecular size, while the interactions are greater than the hydrodynamic interactions. The hydrodynamic diameter of neutral MEEPT remains unchanged with the addition of supporting salts, while for charged MEEPT-BF$_4$, supporting salts decrease the hydrodynamic diameter and enhance interactions. For MEEPT-TFSI, which has a larger solubility compared with MEPPT-BF$_4$, the effects of interactions on the dramatical increase of viscosity at sufficiently high concentration was observed, and the apparent hydrodynamic diameters were calculated in this study, whose change is relative small with the addition of supporting salt. All the solutions are Newtonian, which is explained by comparing the time scale of shear rate to the diffusion-based relaxation time estimate.

It should be noted that the molecular information analysis is based on a continuum assumption, while the sizes of MEEPT and ACN are different by only one order of magnitude. Nonetheless, the hydrodynamic diameters determined from viscometric analysis are comparable to molecular structure dimensions, providing a reasonable estimate. The effect of supporting salts on bulk viscosity is studied, but more complex interactions and conformations of the molecule, such as ion association and liquid solvation, will be the subject of subsequent publications. Other transport properties, such as diffusivity and ionic conductivity versus concentration, remain to be studied, and will enable better prediction of the performance of MEEPT RFBs as well as gaining a better understanding of species association and solvation.

Importantly, the results presented here suggest that MEEPT is a promising ROM candidate with respect to transport properties, as it and its charged cation are "flowable" up to 0.5 M. The result also confirms that the MEEPT/ACN solutions behave like Newtonian liquids over a wide range of shear rates. For molecules with similar size, Newtonian behavior is expected as a result of the relaxation time scale. The model fit and selection method used here also provides a template to analyze measured solution viscosity, from which molecular information can be inferred.

## SUPPLEMENTARY MATERIAL

The viscosity of LiTFSI and TEABF$_4$, the fit procedure of MEEPT and MEEPT-BF$_4$ are shown in supplementary material.

## ACKNOWLEDGMENTS

This work is supported by Joint Center for Energy Storage Research (JCESR), an Energy Innovation Hub funded by the U.S. Department of Energy.

## DATA AVAILABILITY STATEMENT

The data that support the findings of this study are available from the corresponding author upon reasonable request.

# Supporting Information

## A. Experimental Section

### Materials
N-[2-(2-methoxyethoxy)ethyl]phenothiazine (MEEPT, CAS RN 2098786-35-5, >98%) was purchased from TCI America. Hydrochloric acid (ACS grade, 36.5-38%) and silver carbonate ($Ag_2CO_3$, 99%) were purchased from VWR and Sigma Aldrich, respectively. Lithium bis(trifluoromethanesulfonyl)imide (LiTFSI) and tetraethylammonium tetrafluoroborate (TEABF$_4$, 99.9%) were obtained from Biosynth and BASF, respectively. Acetonitrile (ACN, 99.9%) was purchased from Avantor (VWR) and dried using a solvent dispensing system (LC Technology Inc). Diethyl ether (ACS reagent grade) was obtained from Avantor (VWR). Sodium thiosulfate pentahydrate ($Na_2S_2O_3 \cdot 5H_2O$, ≥99%) was obtained from Beantown Chemical (VWR). $^1$H NMR spectra were obtained on 400 MHz Bruker Avance NEO (equipped with a Smart Probe) with 1,4-bis(trifluoromethyl)benzene) (>99%, TCI America) as an internal standard in DMSO-$d_6$ (Cambridge Isotope Laboratories).

### Synthesis of Silver Bis(trifluoromethanesulfonyl)imide (AgTFSI)
AgTFSI was synthesized with some modifications to a reported procedure.[1] LiTFSI (30.0 g, 105 mmol) was dissolved in deionized (DI) water (90 mL) in a 250 mL Erlenmeyer flask containing a stir bar. To this clear colorless solution, conc. hydrochloric acid (10 mL) was added after which the aqueous solution was extracted with diethyl ether (2 x 50 mL). The combined organic layers were concentrated via rotary evaporation, yielding a clear colorless viscous residue. The residue was then dissolved in acetonitrile (500 mL), and silver carbonate (15.9 g, 57.8 mmol) was added. The suspension was stirred for 2 h at room temperature during which a white precipitate formed. The reaction mixture was filtered to remove the solid. The filtrate was then concentrated via rotary evaporation. Diethyl ether (75 mL) was added to the resulting colorless liquid, and the solution was stirred for 1 h after which the solution was concentrated by rotary evaporation. Subsequently DI water (190 mL) was added and the mixture was stirred for another hour. The solution was concentrated by rotary evaporation, which afforded a white solid. The resulting solid was dissolved in 50 mL of diethyl ether, and the solution was filtered to remove undissolved solids. After the solvent was removed via rotary evaporation, the product was obtained as white powder (34.6 g, 85%). $^{13}$C NMR (100 MHz, DMSO-$d_6$) δ 119.5 (q, $J$ = 320 Hz, CF$_3$), which agrees with that reported in literature.[2]

### Synthesis of Radical Cation Salts
N-(2-(2-Methoxyethoxy)ethyl)phenothiazinium tetrafluoroborate (MEEPT-BF$_4$) and N-(2-(2-methoxyethoxy)ethyl)phenothiazinium bis(trifluoromethanesulfonyl)imide (MEEPT-TFSI) were synthesized as previously reported[3] except that AgTFSI synthesized in house was used rather than the commercially available material.

### Solubility Determination of Radical Cation Salts Using $^1$H NMR Spectroscopy
A saturated solution of the compound is made by dissolving an excess amount of solid in electrolyte (ACN, 0.5 M TEABF$_4$/ACN and 0.5 M LiTFSI/ACN). The solution was kept overnight to equilibrate between solute and solvent after which the excess solid was removed by filtering through a syringe filter (25 mm, PTFE). To make an NMR sample a known amount of saturated solution was mixed with a known amount of internal standard (1,4-bis(trifluoromethyl)benzene) at a known concentration in DMSO-$d_6$. When analyzing paramagnetic radical cations, $Na_2S_2O_3$ was added to reduce them radical cations to their neutral form. After making the NMR sample, quantitative $^1$H NMR was recorded using 25 s D1 delay. The solubility of the compound was calculated by integrating the solute and the standard peaks in the NMR spectrum and comparing their ratios.

*Miscibility Determination Procedure*

To determine if MEEPT is miscible in each electrolyte (0.5 M $TEABF_4$/ACN and 0.5 M LiTFSI/ACN), MEEPT was combined with each electrolyte in different weight ratios such as 9:1, 1:1, and 1:9 MEEPT:electrolyte. As MEEPT dissolved completely with electrolyte at all weight ratios, we concluded that MEEPT is miscible with the electrolytes utilized here.

## B. Viscosity of Supporting Salts

In this paper, two supporting salts were studied, $TEABF_4$ and LiTFSI, and the viscosities were tested up to 0.5 M within shear rate of 5,000 to 35,000 $s^{-1}$, as shown in Figure S1. Both the $TEABF_4$/ACN and LiTFSI/ACN solutions show Newtonian behavior, within 5% variation.

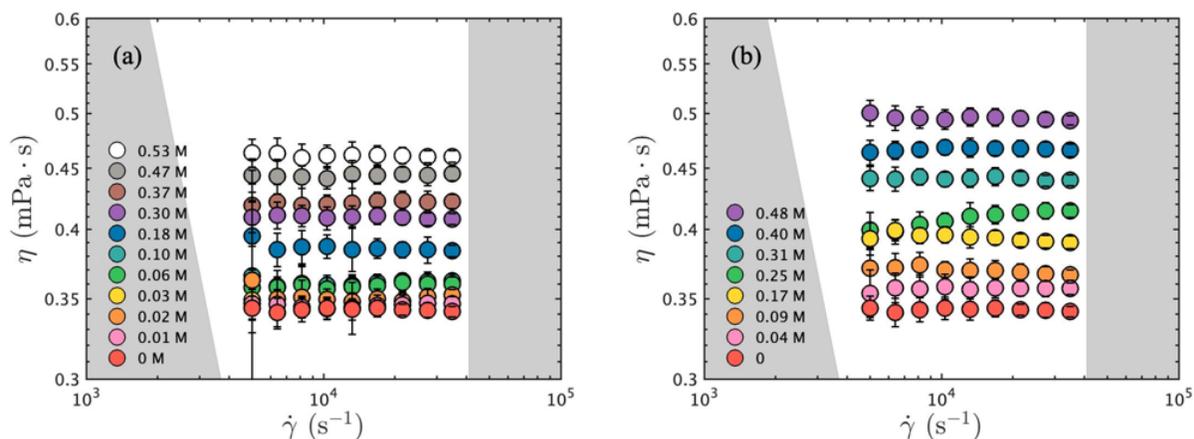

**Figure S1.** The dynamic viscosity of LiTFSI in ACN (a) and for $TEABF_4$ in ACN (b) at 0 – 0.5 M as a function of shear rate. The viscosities show Newtonian behavior with 5% variation.

The average Newtonian viscosities for both supporting salts were obtained by taking an average of nine viscosity data at nine shear rates within 5,000 to 35,000 $s^{-1}$, with the error bars coming from standard deviation and uncertainty propagation, as shown in Figure S2. The orange star points represent the measured average viscosities of $TEABF_4$/ACN solutions at different concentrations of $TEABF_4$, and the green ones represent those of LiTFSI/ACN solutions at different concentrations of LiTFSI.

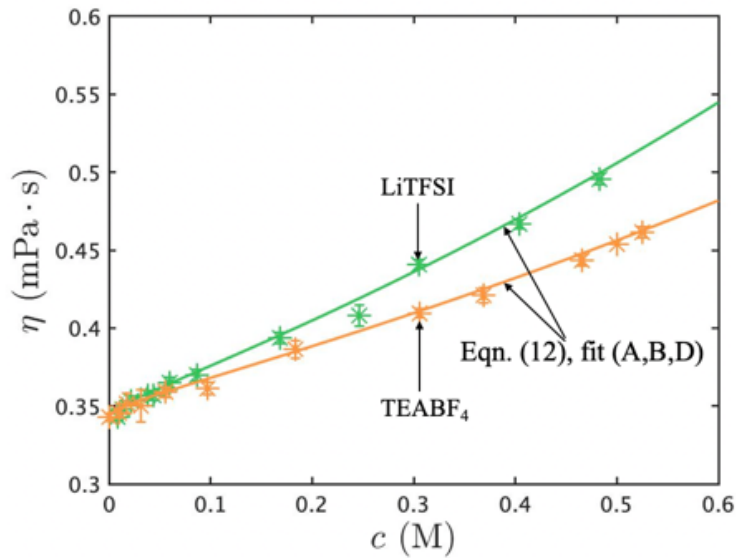

**Figure S2.** The average Newtonian viscosity of TEABF$_4$ in ACN (green) and LiTFSI in ACN (orange) as a function of concentration, using the fitting results of Einstein equation.

The viscosity of the salts in ACN solutions increases faster for LiTFSI than for TEABF$_4$. At a concentration of 0.5 M, the viscosity of TEABF$_4$/ACN solution is ca. 0.45 mPa·s while the LiTFSI/ACN solution is 0.50 mPa·s.

## C. Fit Results of MEEPT/ACN Solutions in Different Supporting Salt Environments

The reduced viscosities of MEEPT/ACN solutions at different concentrations were fit to the reduced Huggins equation Eq. (13) with different numbers of data points, and the most credible fit was found by comparing the BIC of each fit. The fit results are shown in Table S1. From the table, we can see that the most credible fit is the one that uses 7 data points, which corresponds to concentration range of 0.1-0.8 M. This suggests that the Huggins equation is no longer appropriate after the concentration is higher than 0.8 M for MEEPT/ACN solution.

**Table S1.** MEEPT/ACN solution: the fit results of concentration-dependent reduced viscosity to the rearranged Huggins equation (Eq.(13) in the manuscript) with different numbers of data used for fit.

|   | n | $c_{max}$ | BIC | $[\eta]$ | $d_H$ | $k_H$ |
|---|---|---|---|---|---|---|
|   | (-) | (M) | (-) | (L/mol) | (Å) | (-) |
|   | 3 | 0.3 | -17.55 | 0.44 ± 0.06 | 8.26 ± 0.35 | 3.20 ± 0.90 |
|   | 4 | 0.4 | -23.54 | 0.46 ± 0.01 | 8.36 ± 0.02 | 2.61 ± 0.09 |
|   | 5 | 0.5 | -30.39 | 0.47 ± 0.02 | 8.41 ± 0.10 | 2.37 ± 0.22 |
|   | 6 | 0.6 | -35.91 | 0.48 ± 0.02 | 8.47 ± 0.09 | 2.10 ± 0.04 |
| * | 7 | 0.8 | -42.57 | 0.49 ± 0.01 | 8.52 ± 0.07 | 1.95 ± 0.10 |
|   | 8 | 1.0 | -35.10 | 0.45 ± 0.02 | 8.30 ± 0.14 | 2.71 ± 0.21 |
|   | 9 | 1.5 | -18.23 | 0.33 ± 0.06 | 7.51 ± 0.43 | 6.82 ± 1.32 |
|   | 10 | 2.0 | -3.32 | 0.17 ± 0.12 | 6.00 ± 1.36 | 35.13 ± 24.18 |

$c_{max}$ represents the maximum concentration used for the fitting.

\* most credible according to BIC metric, Eq. (14) in the manuscript.

From the most credible fit, the hydrodynamic diameter of MEEPT was determined to be $8.52 \pm 0.07$ Å, while the Huggins coefficient is $1.95 \pm 0.01$.

Similarly, the fit results for MEEPT/ACN solutions with 0.5 M TEABF$_4$ and with 0.5 LiTFSI were calculated and shown below. The most credible fits for both solutions are when $n = 8$ and the corresponding concentration range is 0.1-1.0 M. The fit result are shown in Table S2 and Table S3, with the hydrodynamic diameters of MEEPT being $7.72 \pm 0.12$ and $8.56 \pm 0.16$ Å for two supporting salt environments, and the Huggins coefficients being $4.94 \pm 0.29$ and $2.28 \pm 0.21$ respectively.

**Table S2.** MEEPT/ACN solutions with 0.5 M TEABF$_4$: the fit results of concentration-dependent reduced viscosity to the rearranged Huggins equation (Eq. (13)) with different numbers of data used for fit.

|   | n   | $c_{max}$ | BIC    | $[\eta]$      | $d_H$         | $k_H$         |
|---|-----|-----------|--------|---------------|---------------|---------------|
|   | (-) | (M)       | (-)    | (L/mol)       | (Å)           | (-)           |
|   | 3   | 0.30      | -16.79 | 0.27 ± 0.04   | 6.97 ± 0.38   | 14.90 ± 0.04  |
|   | 4   | 0.40      | -19.43 | 0.30 ± 0.04   | 7.28 ± 0.31   | 9.39 ± 0.04   |
|   | 5   | 0.50      | -23.10 | 0.33 ± 0.03   | 7.51 ± 0.25   | 6.70 ± 1.05   |
|   | 6   | 0.60      | -27.75 | 0.35 ± 0.03   | 7.63 ± 0.21   | 5.57 ± 0.69   |
|   | 7   | 0.80      | -33.85 | 0.36 ± 0.02   | 7.68 ± 0.15   | 5.20 ± 0.42   |
| * | 8   | 1.00      | -40.02 | 0.36 ± 0.02   | 7.72 ± 0.12   | 4.94 ± 0.29   |

$c_{max}$ represents the maximum concentration used for the fit.
\* most credible according to BIC metric, Eq. (14)

**Table S3.** MEEPT/ACN solutions with 0.5 M LiTFSI: the fit results of concentration-dependent reduced viscosity to the rearranged Huggins equation (Eq. (13)) with different numbers of data used for fit.

|   | n   | $c_{max}$ | BIC    | $[\eta]$      | $d_H$         | $k_H$         |
|---|-----|-----------|--------|---------------|---------------|---------------|
|   | (-) | (M)       | (-)    | (L/mol)       | (Å)           | (-)           |
|   | 3   | 0.30      | -16.78 | 0.64 ± 0.04   | 9.30 ± 0.21   | -0.10 ± 0.04  |
|   | 4   | 0.40      | -14.51 | 0.55 ± 0.07   | 8.88 ± 0.39   | 1.29 ± 0.04   |
|   | 5   | 0.50      | -19.35 | 0.56 ± 0.05   | 8.91 ± 0.27   | 1.16 ± 0.45   |
|   | 6   | 0.60      | -22.42 | 0.53 ± 0.04   | 8.77 ± 0.24   | 1.60 ± 0.40   |
|   | 7   | 0.80      | -26.80 | 0.51 ± 0.03   | 8.66 ± 0.19   | 1.97 ± 0.27   |
| * | 8   | 1.00      | -31.63 | 0.49 ± 0.03   | 8.56 ± 0.16   | 2.28 ± 0.21   |

$c_{max}$ represents the maximum concentration used for the fit.
\* most credible according to BIC metric, Eq. (14).

### D. Fit Results of MEEPT-BF$_4$/ACN Solutions

The viscosities of MEEPT-BF$_4$/ACN solutions with no supporting salt (red), with 0.5 M TEABF$_4$ (orange), and with 0.5 M LiTFSI (green) were fit to four different equations: the Einstein, the Huggins, the Jones-Dole, and the extended Jones-Dole equations. The fit results are shown as dotted, dashed, dot-and-dash, and solid lines respectively in Figure S3, Figure S4, and Figure S5, and the details are shown in Table S4, Table S5, and Table S6. From the fit results, we can see that the most credible fits for the three solutions are the extended Jones-Dole equation due to the minimum BIC.

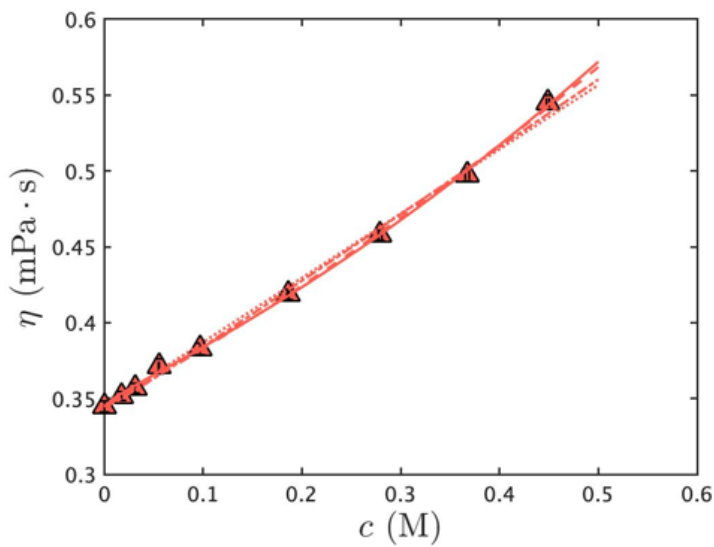

**Figure S3.** MEEPT-BF$_4$/ACN solutions without supporting salt: the fits of concentration-dependent viscosities to the Einstein, the Huggins, the Jones-Dole, and the extended Jones-Dole equations.

**Table S4.** MEEPT-BF$_4$/ACN solutions without supporting salt: the fit results and the BIC of concentration-dependent viscosities fitted to the Einstein, the Huggins, the Jones-Dole, and the extended Jones-Dole equations.

|  | Fit results | BIC |
| --- | --- | --- |
| Einstein equation | $\frac{\eta}{\eta_s} = 1.23c$ | -52.38 |
| Huggins equation | $\frac{\eta}{\eta_s} = 1.06c + 0.48c^2$ | -61.17 |
| Jones-Dole equation | $\frac{\eta}{\eta_s} = -0.06\sqrt{c} + 1.34c$ | -53.86 |
| Extended Jones-Dole equation | $\frac{\eta}{\eta_s} = 0.08\sqrt{c} + 0.79c + 0.82c^2$ | -64.78 |

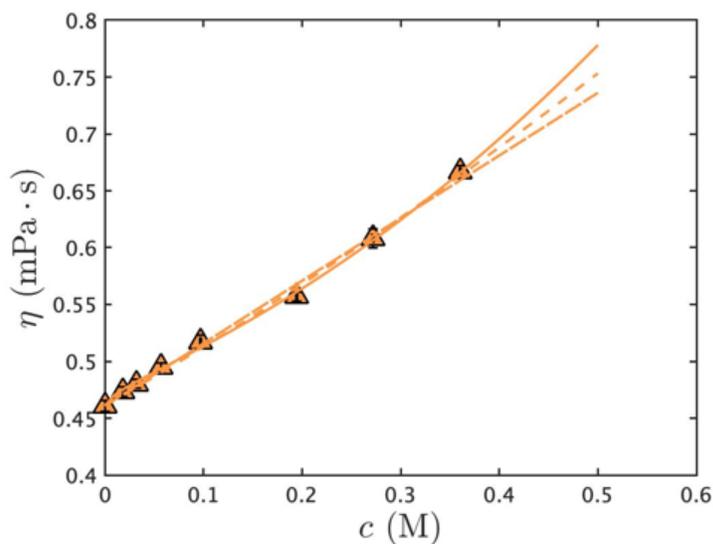

**Figure S4.** MEEPT-BF$_4$/ACN solutions with 0.5 M TEABF$_4$: the fits of concentration-dependent viscosities to the Einstein, the Huggins, the Jones-Dole, and the extended Jones-Dole equations.

**Table S5.** MEEPT-BF$_4$/ACN solutions with 0.5 TEABF$_4$: the fit results and the BIC of concentration-dependent viscosities fitted to the Einstein, the Huggins, the Jones-Dole, and the extended Jones-Dole equations.

|  | Fit results | BIC |
|---|---|---|
| Einstein equation | $\dfrac{\eta}{\eta_s} = 1.19c$ | -48.42 |
| Huggins equation | $\dfrac{\eta}{\eta_s} = 1.09c + 0.36c^2$ | -49.17 |
| Jones-Dole equation | $\dfrac{\eta}{\eta_s} = -0.01\sqrt{c} + 1.20c$ | -47.34 |
| Extended Jones-Dole equation | $\dfrac{\eta}{\eta_s} = 0.16\sqrt{c} + 0.50c + 1.30c^2$ | -56.67 |

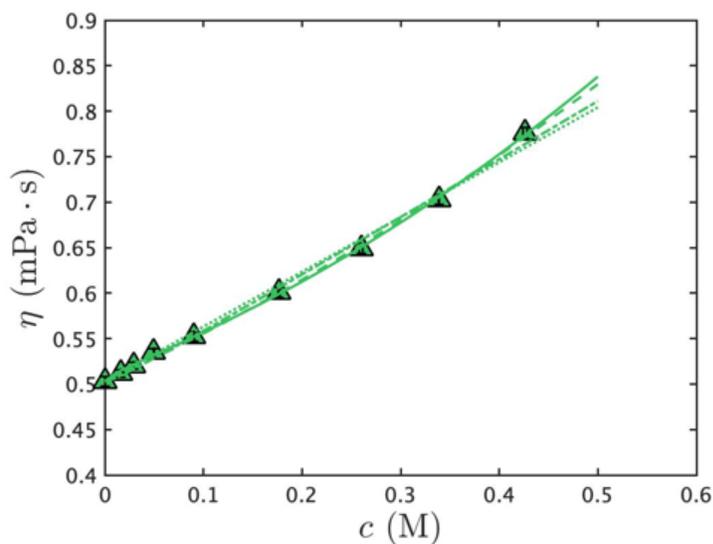

**Figure S5.** MEEPT-BF$_4$/ACN solutions with 0.5 M LiTFSI: the fits of concentration-dependent viscosities to the Einstein, the Huggins, the Jones-Dole, and the extended Jones-Dole equations.

**Table S6.** MEEPT-BF$_4$/ACN solutions with 0.5 M LiTFSI: the fit results and the BIC of concentration-dependent viscosities fitted to the Einstein, the Huggins, the Jones-Dole, and the extended Jones-Dole equations.

|  | Fit results | BIC |
|---|---|---|
| Einstein equation | $\dfrac{\eta}{\eta_s} = 1.20c$ | -50.78 |
| Huggins equation | $\dfrac{\eta}{\eta_s} = 0.99c + 0.63c^2$ | -61.95 |
| Jones-Dole equation | $\dfrac{\eta}{\eta_s} = -0.07\sqrt{c} + 1.32c$ | -52.29 |
| Extended Jones-Dole equation | $\dfrac{\eta}{\eta_s} = 0.10\sqrt{c} + 0.65c + 1.08c^2$ | -69.92 |